\documentclass[11pt,journal,compsoc]{IEEEtran}
\usepackage{cite} 
\usepackage{url} 
\usepackage[utf8]{inputenc} 
\usepackage{booktabs} 
\usepackage{graphicx}
\usepackage[justification=centering]{caption}

\begin{document}
	\title{Mobile Technology in Healthcare Environment: Security Vulnerabilities and Countermeasures}

	\author{
		\IEEEauthorblockN{Sajedul Talukder\IEEEauthorrefmark{1}, Shalisha Witherspoon\IEEEauthorrefmark{2}, Kanishk Srivastava\IEEEauthorrefmark{1}, Ryan Thompson\IEEEauthorrefmark{1}}\\
		\IEEEauthorblockA{\IEEEauthorrefmark{1}Florida International University
			\\\{stalu001, ksriv001, rthom047\}@fiu.edu}\\
		\IEEEauthorblockA{\IEEEauthorrefmark{2}IBM Research
			\\\ swith005@fiu.edu}
	}
	
	\tableofcontents
	\listoffigures
	\listoftables
	\maketitle
	\begin{abstract}
		Mobile devices and technologies offer a tremendous amount of benefits to users, although it is also understood that it introduces a set of challenges when it comes to security, compliance, and risks. More and more healthcare organizations have been seeking to update their outdated technology, and have considered the adoption of mobile devices to meet these needs. However, introducing mobile devices and technology also introduces new risks and threats to the organization. As a test case, we examine Epic Rover, a mobile application that has been identified as a viable solution to manage the electronic medical system. In this paper, we study the insights that the security team needs to investigate, before the adoption of this mobile technology, as well as provide a thorough examination of the vulnerabilities and threats that the use of mobile devices in the healthcare environment brings, and introduce countermeasures and mitigations to reduce the risk while maintaining regulatory compliance. 
		
	\end{abstract}

	\section{Introduction}
	Intensive infiltration of mobile devices into our daily lives has radically changed how we communicate with one another in every sector~\cite{TSRICIEV14,talukder2017usensewer,TSRICEEICT14}. This is also true in the healthcare field, where technology is increasingly playing a role in almost every facet of the industry. The invent of new app technology and digital innovations have now made it possible for consumers to use mobile devices to access patient information, monitor their vital signs, manage and coordinate healthcare, and carry out a wide range of tasks more conveniently. Despite the conveniences brought by the use of mobile devices in a healthcare environment, it also comes with risks like many other sectors~\cite{TalukderC18,talukder2018attacks,TC2017} that must be thoroughly assessed before its adoption. We consider a security team, which consists of the CISO, Security Analyst, Security Engineer, and Chief Compliance Officer, that has been tasked with investigating a viable mobile solution for a hypothetical healthcare organization, and determining whether or not its use is worth the risk to that organization. 
	
	In this paper, we present Epic Rover~\cite{epic}, an innovative mobile application from Epic Systems, which is one of the leaders in healthcare technology systems, and offers a group of reputable mobile apps that have built an integrated platform for almost all areas of healthcare. By utilizing Epic Rover, we will cover the vulnerabilities and risks associated with its use, potential issues with regulatory compliance, industry standards to facilitate compliance, methods for mitigation, and a risk assessment~\cite{kim_solomon_2018} that will determine a recommendation to either adopt or adopt the use of mobile devices for healthcare in the organization.

	\section{Problem Definition and Motivation}
	Sunshine Hospital is a hypothetical large metropolitan organization which is in immediate need to update its population of technological devices. The existing technology infrastructure is fairly old and often fails to meet the current standards of health technology, which includes an abundance of Windows Mobile-based MC75s and legacy windows workstations, most of which face end-of-life by 2020. Moreover, the bulky MC75s and workstations are widely unpopular among the nursing and clinician staff, as most of the old devices have lost their durability and productivity due to excessive wear and tear. The unacceptably slow data transfer rate and poor connectivity have contributed to the rush to look for newer technology, and the urge to adopt mobile devices for day-to-day healthcare activities has been equally heard from the nursing staff and doctors.
	
	We require an app that would allow hospital staff and administrators to provide a more efficient healthcare experience. The app should be able to minimize errors, paper work, and improve efficiency and quality in healthcare management in general. There should be a positive outcome and feedback from the patients as well, as they are the main drivers in our business. Keeping all these factors in mind, our team has worked with the nurses and doctors to find out their expectations and problems with the current technology, allowing us to identify several issues and challenges as we generalized the feedback into a common goal. Furthermore, an equal amount of staff requested a mobile application that was either compatible with an iOS device, or an Android device, depending on their familiarity or level of comfort with the operating system on their own personal devices. 

	\begin{figure}[!h]
		\centering
		\includegraphics[width=0.8\columnwidth]{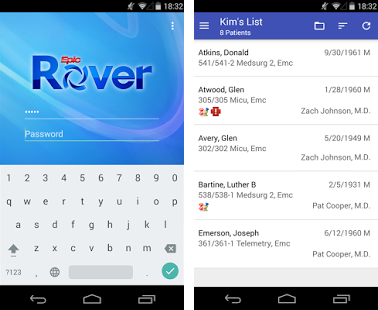} 
		\caption{Epic Rover App}
		\label{fig_sim}
	\end{figure}
	
	
	\section{Proposed Solution -- Epic Rover}
	Considering the requirements mentioned in the previous section, we have identified Epic Rover as the proposed solution to fulfill our company’s needs. Epic Rover~\cite{epic} is a mobile application developed by Epic Systems that facilitates the validation, monitoring, and documentation tasks for doctors and nursing staff. Because of its availability on both iOS and Android, it can satisfy the demand for both groups of staff. Furthermore, it has already been adopted by several large hospitals, including Texas Children’s, Cleveland Clinic, University of Colorado, and Ochsner, making it a more trustworthy solution. 
	\subsection{Basic Design}
	In order to use Epic Rover, an organization must possess a license to Epic Systems 2014 or later. With Rover, hospital staff can use a device mounted barcode scanner to utilize barcoded medication administration (BCMA) features, which allows nurses to positively identify patients, and give them the proper medication by sending alerts if the wrong medication is attempted to be produced~\cite{epic_module}. Furthermore, through the Rover app~\cite{epic_policy}, users have access to Epic’s central data repository, which allows them to collect and review a variety of patient data such as charts, clinical summary of lab reports, allergies, medications, immunizations, medical history, and current condition such as vitals, all in real-time, and by simply tapping the patient’s name on the device. 
	In addition, through Rover, hospital staff can document and update all patient information directly, and send them directly to the central repository. Clinicians collecting patient specimens can use Rover to update collection workflows, print labels, and document the collection, while nurses can update patients’ dosage and medications, and record vitals. Moreover, Rover can find and contact other care providers related to a patient's condition by generating a Care Team Report through secure messaging. These features provide more organization, proficiency, and heightened communication between nurses and other healthcare providers, ultimately improving patient safety.
	\subsection{Basic security mechanisms}
	The Epic Rover mobile application implements several basic mechanisms to ensure that electronic health records are securely accessed only by authorized users, and protect the confidentiality, integrity, and availability of patients’ personal information. Table~\ref{tab:basicsec} depicts Epic Rover’s security controls, along with the threats and vulnerabilities that they mitigate:
	
	\begin{table}[!htbp]
		\centering
		\begin{tabular}{|p{8em}|p{10em}|p{10em}|}
			\toprule
			\textbf{Threats} & \textbf{Vulnerabilities} & \textbf{Security Control} \\
			\midrule
			Man-in-the-middle attack & Unencrypted data & TLS/SSL: Supports Transport Layer Security (TLS)/Secure Sockets Layer (SSL) and encryption for communications to ensure that all data is transmitted securely over HTTPS.  \\
			\midrule
			Unauthorized access & Lack of proper access controls & Two-factor authentication: Two-factor authentication is embedded into the app, providing stronger access controls. Additionally, it is not possible to access the app from someone else’s device, as Epic Rover is assigned to specific devices only. This ensures that, even if someone’s credentials are stolen, their account will still be secured as long as they still possess the actual device. \\
			\midrule
			Malicious application sharing data with Epic Rover & Improper implementation of application verification & Signature-based permissions: Ensures that the apps accessing the data among themselves are signed using the same signing key, thus offering a more streamlined and secure user experience. \\
			\midrule
			Code injection attack & Improper input validation & No dynamic code loading: Epic Rover completely runs over native code, meaning it does not load code from outside of the application environment.  \\
			\midrule
			Violation of patient’s private or confidential information & Inadequate review of privacy policies & Data privacy: Epic Rover states that it does not sell or license any information that it may collect from the user or provider, nor does it store any personal information on the device, or send directly to Epic.  \\
			\bottomrule
		\end{tabular}%
		\vspace{1em}
		\caption{Basic security mechanisms}
		\label{tab:basicsec}%
	\end{table}%
	
	\subsection{Regulatory Compliance}
	While the adoption of mobile devices for health care applications and management systems could offer a more convenient experience for our staff, it is important to be aware that it also introduces a new set of risks and challenges, particularly in regards to adhering to Federal Government regulations. As professionals in the Health Care Industry, we are listed as covered entities under the U.S. Department of Health \& Human Services' Health Insurance Portability and Accountability Act of 1996 (HIPAA)~\cite{HIPAA_guidance}. By electing to use mobile technologies, we would be required to comply with HIPAA’s Security Rule, which mandates a set of standards for covered entities to follow in order to secure the confidentiality, integrity, and availability of Electronic Protected Health Information (EPHI).The standards developed in HIPAA’s Security Rule are divided into three sections: Administrative Safeguards, Physical Safeguards, and Technical Safeguards.  While the implementations for some standards are required for compliance, others are merely addressable, leaving it up to the organization to determine whether or not it is appropriate to adopt based on their needs. Therefore, it is crucial to have a general understanding and familiarity with the Security Rule standards, so that our organization not only remains compliant, but can also use the standards as a baseline to keep our clients' invaluable information protected, and maintain our reputation as a trustworthy and security-conscious company. 
	
	\textbf{Administrative Safeguards (164.308)}
	The Administrative Safeguards for the HIPAA Security Rule sets forth a list of security measures related to administrative actions, policies, and procedures, to ensure protection of EPHI. There are a total of nine standards that include security management process, assigned security responsibility, workforce security, information access management, security awareness and training, security incident procedures, contingency plan, evaluation, business associate contracts and other arrangement.\\
	\textbf{Physical Safeguards (164.310)}
	The Physical Safeguards for the HIPAA Security Rule sets forth a list of applicable policies and physical measures to protect EPHI from threats such as unauthorized access, and natural disasters. There are a total of four standards that include facility access controls, workstation use, workstation security, device and media controls.\\
	\textbf{Technical Safeguards (164.312)}
	The Technical Safeguards for the HIPAA Security Rule sets forth a list of security measures related to the use of technology to protect EPHI, and such policies and procedures implemented for access control. There are a total of five standards that include access control, audit controls, integrity, person or entity authentication and transmission security.
	
	Although there are numerous standards to comply with in HIPAA, they provide a straightforward 
	
	\begin{table*}[]
		\centering
		\resizebox{0.97\textwidth}{!}{%
			\begin{tabular}{|r|r|r|r|}
				\toprule
				\multicolumn{1}{|p{5em}|}{\textbf{Function}} & \multicolumn{1}{p{16em}|}{\textbf{Category}} & \multicolumn{1}{p{14em}|}{\textbf{Subcategory}} & \multicolumn{1}{p{34em}|}{\textbf{HIPAA Control Mapping}} \\
				\midrule
				\multicolumn{1}{|p{5em}|}{\textbf{Identify }} & \multicolumn{1}{p{16em}|}{Risk Management: The organization’s priorities, constraints, risk tolerance, and assumptions are established and used to support operational risk decisions.  } & \multicolumn{1}{p{14em}|}{Risk management processes are established, managed, and agreed to by organization stakeholders} & \multicolumn{1}{p{34em}|}{164.308(a)(1)(ii)(B) – Risk Management} \\
				\cmidrule{3-4}          &       & \multicolumn{1}{p{14em}|}{Organizational risk tolerance is determined and clearly expressed} & \multicolumn{1}{p{34em}|}{164.308(a)(1)(ii)(B) – Risk Management} \\
				\cmidrule{3-4}          &       & \multicolumn{1}{p{14em}|}{The organization’s determination of risk tolerance is informed by its role in critical infrastructure and sector specific risk analysis} & \multicolumn{1}{p{34em}|}{164.308(a)(1)(ii)(B) – Risk Management, 164.308(a)(6)} \\
				&       &       & \multicolumn{1}{p{14em}|}{(ii) – Response and Reporting, } \\
				&       &       & \multicolumn{1}{p{34em}|}{164.310(a)(2)(i) – Contingency operations} \\
				&       &       &  \\
				\midrule
				\multicolumn{1}{|p{5em}|}{\textbf{Protect}} & \multicolumn{1}{p{16em}|}{Protective Technology: Technical security solutions are managed to ensure the security and resilience of systems and assets, consistent with related policies, procedures, and agreements} & \multicolumn{1}{p{14em}|}{Audit/log records are determined, documented, implemented, and reviewed in accordance with policy} & \multicolumn{1}{p{34em}|}{164.308(a)(1)(ii)(D) –Information system activity review, 164.308(a)(5)(ii)(C) – Log-in monitoring, } \\
				&       &       & \multicolumn{1}{p{34em}|}{164.310(a)(2)(iv) – Maintenance records, 164.310(d)(2)(iii) - Accountability, } \\
				&       &       & \multicolumn{1}{p{34em}|}{164.312(b) – Audit controls} \\
				&       &       &  \\
				\cmidrule{3-4}          &       & \multicolumn{1}{p{14em}|}{Removable media is protected and its use restricted according to policy} & \multicolumn{1}{p{34em}|}{164.308(a)(3)(i) – Workforce security, 164.308(a)(3)(ii)(A) – Authorization/Supervision, } \\
				&       &       & \multicolumn{1}{p{34em}|}{164.310(d)(1) – Device and media controls, 164.312(a)(1) – Access control, } \\
				&       &       & \multicolumn{1}{p{34em}|}{164.312(a)(2)(iv) – Encryption/Decryption, 164.312(b) – Audit controls} \\
				&       &       &  \\
				\cmidrule{3-4}          &       & \multicolumn{1}{p{14em}|}{Access to systems and assets is controlled, incorporating the principle of least functionality} & \multicolumn{1}{p{34em}|}{164.308(a)(3) – Assigned security responsibility, 164.308(a)(4) – Information access management, } \\
				&       &       & \multicolumn{1}{p{34em}|}{164.310(a)(2)(iii) – Access control and validation procedures, 164.310(b) – Workstation use, 164.310(c) – Workstation security, } \\
				&       &       &  \\
				\cmidrule{3-4}          &       & \multicolumn{1}{p{14em}|}{Communications and control networks are protected} & \multicolumn{1}{p{34em}|}{164.308(a)(1)(ii)(D) – Information system activity review, 164.312(a)(1) – Access control, } \\
				&       &       & \multicolumn{1}{p{34em}|}{164.312(b) – Audit controls, 164.312(e) – Transmission security} \\
				&       &       &  \\
				\midrule
				\multicolumn{1}{|p{5em}|}{\textbf{Detect}} & \multicolumn{1}{p{16em}|}{Anomalies and Events: Anomalous activity is detected in a timely manner and the potential impact is understood} & \multicolumn{1}{p{14em}|}{A baseline of network operations and expected data flows for users and systems is established and managed} & \multicolumn{1}{p{34em}|}{164.308(a)(1)(ii)(D) – Information system activity review,} \\
				&       &       & \multicolumn{1}{p{34em}|}{ 164.312(b) – Audit controls} \\
				&       &       &  \\
				\cmidrule{3-4}          &       & \multicolumn{1}{p{14em}|}{Detected events are analyzed to understand attack targets and methods} & \multicolumn{1}{p{34em}|}{164.308(6)(i) – Security incident procedures} \\
				\cmidrule{3-4}          &       & \multicolumn{1}{p{14em}|}{Impact of events is determined} & \multicolumn{1}{p{34em}|}{164.308(a)(6)(ii) – Response and Reporting} \\
				&       &       &  \\
				\midrule
				\multicolumn{1}{|p{5em}|}{\textbf{Respond}} & \multicolumn{1}{p{16em}|}{Mitigation: Activities are promptly performed to prevent further expansion of the event, mitigate the effects caused by the incident, and eradicate it. } & \multicolumn{1}{p{14em}|}{Incidents are contained} & \multicolumn{1}{p{34em}|}{164.308(a)(6)(ii) – Response and Reporting} \\
				\cmidrule{3-4}          &       & \multicolumn{1}{p{14em}|}{Incidents are mitigated} & \multicolumn{1}{p{34em}|}{164.308(a)(6)(ii) – Response and Reporting} \\
				\cmidrule{3-4}          &       & \multicolumn{1}{p{14em}|}{Newly identified vulnerabilities are mitigated or documented as accepted risks} & \multicolumn{1}{p{34em}|}{164.308(a)(1)(ii)(A) – Risk analysis 164.308(a)(1)(ii)(B) – Risk management , } \\
				&       &       &  \\
				\midrule
				\multicolumn{1}{|p{5em}|}{\textbf{Recover}} & \multicolumn{1}{p{16em}|}{Communications: Restoration activities are coordinated with internal and external parties} & \multicolumn{1}{p{14em}|}{Public relations are managed} & \multicolumn{1}{p{34em}|}{164.308(a)(6)} \\
				&       &       & \multicolumn{1}{p{14em}|}{(i) – Security Incident Procedures} \\
				&       &       &  \\
				\cmidrule{3-4}          &       & \multicolumn{1}{p{14em}|}{Reputation after an event is repaired} & \multicolumn{1}{p{34em}|}{164.308(a)(6)} \\
				&       &       & \multicolumn{1}{p{14em}|}{(i) – Security Incident Procedures} \\
				&       &       &  \\
				\cmidrule{3-4}          &       & \multicolumn{1}{p{14em}|}{Recovery activities are communicated to internal stakeholders and executive and management teams} & \multicolumn{1}{p{34em}|}{164.308(a)(6)(ii) – Response and reporting, 164.308(a)(7)(ii)(B) – Disaster Recovery Plan, } \\
				&       &       &  \\
				\bottomrule
			\end{tabular}%
		}
		\vspace{1em}
		\caption{Correlation of administrative, physical, and technical safeguard in the HIPAA Security Rule to a function from the NIST Cybersecurity Framework.}
		\label{tab:ocr}%
	\end{table*}%

	and effective guideline to helping ensure the confidentiality, integrity, and availability of EPHI, and should be taken seriously. Failure to comply with HIPAA may result in criminal penalties, as specified in the HIPAA Enforcement Rule.
	\subsection{Industry Standards -- NIST}
	Safeguarding the confidentiality, integrity, and availability of EPHI, while also making sure to comply with HIPAA’s Security Rule can be challenging, especially with the inherited risks that the use of mobile technology introduces. Fortunately, there are industry standards available that can be used to facilitate HIPAA compliance, and improve the overall security measures that we have already developed and implemented. While there are a number of standards formulated by different organizations to choose from, the standard we have selected as a guideline to fit our organizations’ needs is from the National Institute of Standards and Technology (NIST)~\cite{NIST_HIPAA}, which is under the U.S. Department of Commerce. In response to President Obama’s Executive Order for Improving Critical Infrastructure Cybersecurity, NIST developed a Cybersecurity Framework~\cite{cybersecurity_framework} that outlines a set of standards, guidelines, and best practices to assist organizations in managing and controlling the risks and threats to cybersecurity. Identified as valuable guidance to improve security programs and aid in HIPAA compliance, the Office of Civil Rights (OCR), which is responsible for auditing and enforcing HIPAA compliance, developed a crosswalk that creates a mapping between the standards listed in the NIST Cybersecurity Framework, and those found in the HIPAA Security Rule. This makes it easier to identify how the Cybersecurity Framework compliments the Security Rule, and also identify gaps in security measures that may not have been met following either standard alone, allowing a more comprehensive and enhanced safeguarding for our organization’s EPHI. 
	
	One of the key components of the NIST Cybersecurity Framework is the Framework Core, which are comprised of five functions that are essential to cybersecurity management. Each function is divided into categories, which are the desired outcomes of security measures associated with the function. The categories can be further divided into subcategories, which further detail and specify the outcomes related to the security control function. Lastly, each function has an informative reference, which map to sections of existing standards to “illustrate” ways that the outcomes of each function can be implemented. The five functions are listed below:
	\begin{itemize}
		\item \textbf{Identify:} Understanding risks and threats to cybersecurity in order to implement appropriate policies and procedures to mitigate them.
		\item \textbf{Protect:} Providing safeguards to prevent systems and assets from being compromised.
		\item \textbf{Detect:} Implementing methods to positively identify when a security event has taken place.
		\item \textbf{Respond:} Executing actions as part of a plan to respond effectively to an identified security event.
		\item \textbf{Recover:} Executing actions as part of a plan to resume normal business operations following a security event.
	\end{itemize}
	
	\begin{table}[]
		\centering
		\begin{tabular}{|p{6em}|p{14em}|p{8em}|}
			\toprule
			\textbf{Vulnerability} & \textbf{Description} & \textbf{Countermeasure} \\
			\midrule
			\textbf{Outdated Software Version and Delay in Patching:} & Not patching the OS or software on a regular basis leaves the system in a vulnerable state, and could allow new and identified threats to exploit the system’s lack of updates. This increases the system’s susceptibility to malware, and can result in the data being compromised. & Automatic updates, policies requiring patching when made available \\
			\midrule
			\textbf{Insufficient Authorization} & Authorization procedures should be specifically defined for users in respect to their role in the organization, their status, and their department, to avoid users who aren’t authorized to view personal data gaining access to it. & Two-factor authentication, access control lists, IDS/IPS \\
			\midrule
			\textbf{Improper use of Device} & If a user carries out unacceptable behavior on a device containing EPHI, such as accessing untrustworthy sites, downloading media, or emailing personal information, they may not only put EPHI at risk, but also the network. & Acceptable Use Policy, system logs, training \\
			\midrule
			\textbf{Connection to Unsecure or Un-trusted Network} & If there is an incoming and outgoing of data over a network which has not been secured or verified using a trusted certificate, the data is exposed to invalid access and modification, especially without the use of cryptography. & Restrict devices to intranet connection, VPN, firewall, encryption \\
			\midrule
			\textbf{Jailbreaking} & Rooting or jailbreaking a mobile device may leave it open to malicious attacks, as the encryption protection gets bypassed if the app is running on a rooted device. & Acceptable Use Policy, perform regular system test/analysis \\
			\midrule
			\textbf{Unattended Device} & If a device containing EPHI is not properly monitored, it may be accessed or stolen by unauthorized users, exposing personal and confidential data.  & Remote wiping, lock inactive devices, monitoring (cameras and logs) \\
			\bottomrule
		\end{tabular}%
		\vspace{1em}
		\caption{Vulnerabilities and Countermeasures}
		\label{tab:vulnerabilities}%
	\end{table}%
	
	Table~\ref{tab:ocr} shows OCR’s mapping that demonstrates how each administrative, physical, and technical safeguard in the HIPAA Security Rule correlates to a function from the NIST Cybersecurity Framework.
	
	Following the standards, guidelines, and best practices introduced in the NIST Cybersecurity Framework should have a positive impact on our overall security program, and help us to ensure we remain HIPAA compliant as we take into consideration the adoption of mobile devices and technology, and the additional risks that doing so may bring.

	\subsection{Threats and Attacks}
	As the use of mobile devices in the healthcare environment is on a continuous rise, so too are the threats against them. There are four categories of malicious attacks to security, which includes interruptions, interceptions, modifications, and fabrications~\cite{CBR_2015,health_privacy}. Below are some of the top identified threats from each category of attack, and how they can negatively affect the security of EPHI.\\ 
	\textbf{Mobile Ransomware (Interruption):} Mobile Ransomware can `lock out' patient information contained in the device, and then demand a ransom in exchange for restoring access to the data and its availability, usually in the form of Bitcoin to avoid tracking, which affects the availability of systems.\\
	\textbf{Mobile Spyware (Interception):} Mobile spyware is a program that unknowingly gets loaded onto a mobile device and records critical user information, and affects the confidentiality. Having a healthcare app on the same infected device can result in the spyware picking up the login credentials for the app and as a result, unauthorized access to EPHI.\\
	\textbf{Compromised Servers (Modification):} As a host to Epic’s central data repository, servers can be a prime target for attackers, who may not only seek to access confidential information, but also modify or delete information contained in the database, affecting the integrity of the EPHI.\\
	\textbf{Social Engineering (Fabrication):} With personal and valuable information contained in a single location, mobile devices containing EPHI would be an attractive target for attackers such as social engineers, who may manipulate unknowing individuals into willingly providing them with access to EPHI, and giving them the opportunity to steal the device itself, making it a serious threat to the assurance of CIA. Examples include phishing, or vishing. 
	
	\section{Vulnerabilities}
	Although we have identified common threats, it is only through vulnerabilities that these threats are able to be exploited. According to a security report from Maryland based IT security provider, Arxan Technologies, at least two of the top 10 OWASP (Open Web Application Security Project) vulnerabilities were present in the majority of the mobile health apps that they tested, despite nearly 80 percent of these applications having been approved by the FDA. Thus, we have identified important vulnerabilities that the use of mobile devices in a healthcare environment exposes our organization to, as well as appropriate countermeasures to mitigate them, as depicted in the table~\ref{tab:vulnerabilities}: 
	
	\section{Risk Mitigation and Management}
	After evaluating the threats and vulnerabilities to EPHI, it is imperative to develop and adopt appropriate mitigations to reduce risks and maintain compliance. The following are appropriate mitigations to manage risks, which were recommended by the Department of Health and Human Services for guidance, and involve managing the risks associated with storing, accessing, and transmitting EPHI.
	\subsection{User Training and Awareness:} Users are identified as the weakest link in security, which increases the risk of a situation leading to compromised EPHI. Therefore, it is essential that the workforce be trained and given clear and concise instructions on the steps that need to be taken in order to follow best practices to avoid any risk to EPHI. This includes:
	\begin{itemize}
		\item Implementing policies to hold regular information and training sessions regarding acceptable use of mobile devices, careful monitoring of devices, and having it enforced. 
		\item Developing password management procedures for changing and safeguarding passwords.
		\item Maintaining system logs for accountability, and deterring improper use.
	\end{itemize}
	
	\begin{table}[t]
		\centering
		\resizebox{0.99\textwidth}{!}{%
			\begin{tabular}{|p{5em}|p{4em}|p{4em}|p{4em}|p{8em}|p{8em}|p{6em}|p{6em}|p{8em}|}
				\toprule
				\multicolumn{4}{|r|}{}        & \multicolumn{3}{p{22em}|}{\textbf{Mitigating Controls}} & \multicolumn{2}{r|}{} \\
				\midrule
				\textbf{Potential Threats  and Vulnerabilities} & \textbf{Probability of Occurrence (H, M, L)} & \textbf{Potential Impact/  Severity   (H, M, L)} & \textbf{Inherent Risk Rating       (H, M, L)} & \textbf{Administrative} & \textbf{Technical} & \textbf{Physical Security} & \textbf{Residual Risk} & \textbf{Comments} \\
				\midrule
				Mobile Malware: Trojans and Viruses & H     & H     & H     & Implement policies requiring anti-virus on all mobile devices containing EPHI  & Install a firewall and antivirus software on all devices on  network & Monitor devices & M (antivirus must be kept up to date) & Regularly update anti-malware software  \\
				\midrule
				Zero-day Vulnerabilities & L     & H     & M     & Implement policies and procedures for regularly testing application & Stay up to date on applying patches  & IDS/IPS  & M (New risks will always come) & N/A \\
				\midrule
				Log on credentials lost or stolen & M     & H     & M     & Implement policies and procedures for strong passwords, and two-factor authentication  & Implement access controls for credentials to expire after a certain amount of time, and requiring long characters with special characters & Lock out device  when not in use, use biometrics for authentication and access   & M (social engineering can bypass security measures) & Immediately reset passwords that are stolen, or remove permission for users who log-on info is compromised \\
				\midrule
				Brute force attacks & M     & H     & H     & Implement policies to change password every 30 days, and two-factor authentication  & Lock at accounts after a certain number of failed login attempts & Lock unattended devices and store in secure location & L (biometrics should decrease risk) & Use logs to monitor log-in attempts \\
				\midrule
				Data modified during transmission  & M     & H     & M     & Implement policies to allow devices containing EPHI to only connect to the intranet, and prohibit access to public networks & Transmission through secure channels such as SSL/TSL over HTTPS  & Secure wireless access points  & M     & Set up VPN if remote access is necessary, or allowed \\
				\midrule
				Lost or stolen device & M     & H     & H     & Implement policies requiring encryption for all devices containing EPHI & Allow remote wiping of data for lost or stolen devices  & Store devices in secure location, and never leave unattended  & M (hospital staff may be careless)  & Hold regular training for user awareness \\
				\midrule
				Operating System or Application on Mobile device is outdated & H     & H     & H     & Implement policies requiring updates  within 24 hours after notification   & Enable automatic updates  & Leave notification and reminders  & M (The application must be updated ) & Train users, and enforce compliance \\
				\bottomrule
			\end{tabular}%
		}
		\vspace{1em}
		\caption{Qualitative Risk Assessment Matrix}
		\label{tab:qualitative risk assessment matrix}%
	\end{table}%

	\clearpage
	
	\begin{itemize}
		\item Keeping firmware and apps updated, and applying patches when immediately available.
		\item Restricting apps on mobile devices that relate to transmission of EPHI over a secure and private network. For example, emails should be sent only over the organization’s private email server, and use of apps like Dropbox and Google drive should be discouraged.
	\end{itemize}
	
	\subsection{Risk Management for Accessing EPHI}
	\begin{itemize}
		\item Implement Two-factor authentication to restrict unauthorized access.
		\item Implement access controls to categorize users on the basis of their job function in order to restrict access to EPHI to only authorized users.
		\item Install and update antivirus protection regularly in order to create a secure environment for accessing the data.
		\item Install firewalls to filter traffic on the network.
	\end{itemize}
	
	\subsection{Risk Management for Storing EPHI}
	\begin{itemize}
		\item Utilize encryption on mobile devices containing EPHI to protect the confidentiality and integrity.
		\item Implement locking methods for unattended or inactive devices.
		\item Maintain backups of EPHI contained on devices.
		\item Have methods to log activity for accountability.
		\item Implement procedures to remotely wipe data contained on lost or stolen devices.
		
	\end{itemize}
	\subsection{Risk Management for Transmitting EPHI}
	\begin{itemize}
		\item Mobile devices containing EPHI should only connect to the organization’s intranet, and never connect to public access points.
		\item Use of non-secure transmission modes such as non-organizational email systems should be prohibited.
		\item Only the use of secure connections for transmission such as SSL, and the use of message-level standards such as S/MIME, SET, PEM, PGP, should be allowed.
	\end{itemize}
	
	Identifying the risk and applying the appropriate countermeasures and mitigations is an important step in determining how the use of mobile of devices affects our organization’s security, and a crucial factor for the final demonstration in our report: the risk assessment. 
	
	\section{Risk Assessment and Recommendation}
	Table~\ref{tab:qualitative risk assessment matrix} shows the qualitative risk assessment matrix developed by taking into account potential threats and vulnerabilities, the probability of their occurrence, the potential impact and severity if they were to occur, and the inherent risk by introducing mobile devices. The results of each are rated either (H)igh, (M)edium, or (L)ow. Additionally, administrative, technical, and physical mitigating controls for the risk are also identified, along with the residual risk after applying the controls.

	When considering the initial risks and the residual risk levels after implementing security management measures, it is expected that the mitigation controls should reduce the risks to appropriate levels. Thus, after properly identifying and qualifying the potential vulnerabilities and threats, and thoroughly examining the proper countermeasures and safeguards as well, it is our recommendation that our organization adopt the use of mobile devices, as our security team is prepared to address the challenges that the use of mobile devices would bring, and believe that the benefits derived from using the Epic Rover mobile app, along with its security mechanisms already set in place, outweigh the risks. 
	
	\section{Conclusions}
	Despite having some security vulnerabilities, deploying Epic Rover with adequate countermeasures will simplify the daily operations of healthcare management and tasks for the healthcare staff, while providing safeguards to ensure the confidentiality, integrity, and availability of the EPHI contained on the mobile device. Taking measures to ensure compliance by following the NIST Cybersecurity Framework should ease the process of introducing mobile devices into our environment, and guide us in developing and implementing the proper policies and procedures to further safeguard the EPHI that we create, store, access, and transmit. We recommend that our next steps be in determining the appropriate business model for using mobile devices, such as Bring-Your-Own-Device (BYOD), or Corporate-Owned, Personally-Enabled, and implement the safeguards recommended in the risks assessment to ensure a smooth and secure adoption.

	
	\clearpage
	\bibliographystyle{IEEEtran}
	\bibliography{references,talukder}
	

\end{document}